\documentclass[%
reprint,
superscriptaddress,
aps,
prl,
portrait,
]{revtex4-2}
\usepackage{blindtext}
\usepackage{amsmath}
\usepackage{bm}
\usepackage{xcolor}
\usepackage{soul}
\usepackage{graphicx}
\usepackage{subfig}
\usepackage[normalem]{ulem}
\usepackage{siunitx}
\usepackage[version=4]{mhchem} 
\usepackage{caption}
\usepackage{pdfpages}
\graphicspath{ {./figures/} }

\makeatletter
\AtBeginDocument{\let\LS@rot\@undefined}
\makeatother

\begin{document}

\title{How charges separate when surfaces are dewetted}

\author{Aaron D. Ratschow}
\affiliation{
 Institute for Nano- and Microfluidics, TU Darmstadt,\\
 Alarich-Weiss-Stra{\ss}e 10, D-64237 Darmstadt, Germany
}
\author{Lisa S. Bauer}
\affiliation{
 Institute for Nano- and Microfluidics, TU Darmstadt,\\
 Alarich-Weiss-Stra{\ss}e 10, D-64237 Darmstadt, Germany
}
\author{Pravash Bista}
\affiliation{
 Max Planck Institute for Polymer Research, \\
 Ackermannweg 10, 55128 Mainz, Germany
}
\author{Stefan A. L. Weber}
\affiliation{
 Max Planck Institute for Polymer Research, \\
 Ackermannweg 10, 55128 Mainz, Germany
}
\affiliation{
 Department of Physics, Johannes Gutenberg University, \\
 Staudingerweg 10, 55128 Mainz, Germany
}
\author{Hans-Jürgen Butt}
\affiliation{
 Max Planck Institute for Polymer Research, \\
 Ackermannweg 10, 55128 Mainz, Germany
}
\author{Steffen Hardt}
\email{hardt@nmf.tu-darmstadt.de}
\affiliation{
 Institute for Nano- and Microfluidics, TU Darmstadt,\\
 Alarich-Weiss-Stra{\ss}e 10, D-64237 Darmstadt, Germany
}

\date{\today}

\begin{abstract}
Charge separation at moving three-phase contact lines is observed in nature as well as technological processes. Despite the growing number of experimental investigations in recent years, the physical mechanism behind the charging remains obscure. Here we identify the origin of charge separation as the dewetting of the bound surface charge within the electric double layer by the receding contact line. This charge depends strongly on the local electric double layer structure close to the contact line, which is affected by the gas-liquid interface and the internal flow of the liquid. We summarize the charge separation mechanism in an analytical model that captures parametric dependencies in agreement with our experiments and numerical simulations. Charge separation increases with increasing contact angle and decreases with increasing dewetting velocity. Our findings reveal the universal mechanism of charge separation at receding contact lines, relevant to many dynamic wetting scenarios, and provide a theoretical foundation for both fundamental questions, like contact angle hysteresis, and practical applications. 
\end{abstract}

\maketitle

Liquid drops interacting with solid surfaces play a role in many natural and technological processes. In nature, organisms have developed surfaces from which drops easily roll off to prevent fouling \cite{Barthlott.1997} or surfaces for fog harvesting \cite{Parker.2001}. Technological applications relying on the interaction between drops and surfaces include inkjet printing \cite{Lohse.2022}, condensation heat transfer \cite{Hu.2021,Zhang.2022}, open droplet microfluidics \cite{Pollack.2000,Cho.2003}, and application of spray droplets to plant leaves \cite{Kovalchuk.2021}, among others. In many of these scenarios, dynamic wetting plays a key role as a solid-liquid-gas three-phase contact line gets displaced along a surface. 

Already decades ago, it was noticed that water drops sliding along a surface acquire a charge  \cite{Yatsuzuka.1994}, but only recently this phenomenon has moved into the focus of intense research activities \cite{Lin.2014,Sun.2015,Boamah.2019,He.2019,Xu.2020,Wu.2020,Shahzad.2018,Stetten.2019,Sun.2019,Lin.2020,Choi.2013,Nauruzbayeva.2020,Poli.2020}.
Slide electrification can either be a desired or an undesired phenomenon. In semiconductor manufacturing, wafers get damaged by electrostatic discharges that occur when rinsing them with aqueous solutions \cite{Dhane.2011,Sano.2016,Guo.2012}. On the other hand, the charge a drop acquires can be employed for energy harvesting \cite{Helseth.2016,Shahzad.2018,Sun.2015,Wang.2020,Xu.2020,Wu.2020}. Slide electrification also occurs in nature when a water drop hits a plant leaf \cite{Armiento.2022} but is suppressed by conductive substrates \cite{Li.2022}. Recently, experiments with drops sliding along a number of different surfaces have demonstrated the dramatic influence the charging has on the motion of drops \cite{Li.2022}. It was shown that the electrostatic interaction between a drop and the surface charge it leaves behind is comparable in magnitude to the friction force it would experience in the absence of electrostatic interactions, and that it even reduces the contact angle. Thus, drop charging is of fundamental importance in many dynamic wetting scenarios and profoundly influences drop trajectories along solid surfaces.

In spite of the widespread importance of drop charging, the underlying physical mechanisms of charge separation between the liquid and the solid surface have remained obscure. The phenomenon has commonly been attributed to ionic charges \cite{Kudin.2008,Zimmermann.2009,Sosa.2022}. Models based on the surface chemistry have been proposed and fitted to data that however neglect processes close to the contact line \cite{Sosa.2022,Helseth.2023}. Recently, a charge separation mechanism based on additional electron transfer was suggested \cite{Lin.2020,Zhan.2020}. Yet, no theory to which experimental data could be compared has been elaborated. In the present paper, we present a theory that explains the charge separation mechanism based on transport processes in the vicinity of a receding contact line and compare theoretical predictions to experimental data. The theory is not only applicable to drops sliding along surfaces, but also to other processes in which receding contact lines play a role, such as liquid film dynamics \cite{Thiele.1998,Eggers.2004,Martin.1998,Edwards.2016,Mulji.2010}.

\begin{figure}
  \includegraphics[width=0.99\columnwidth]{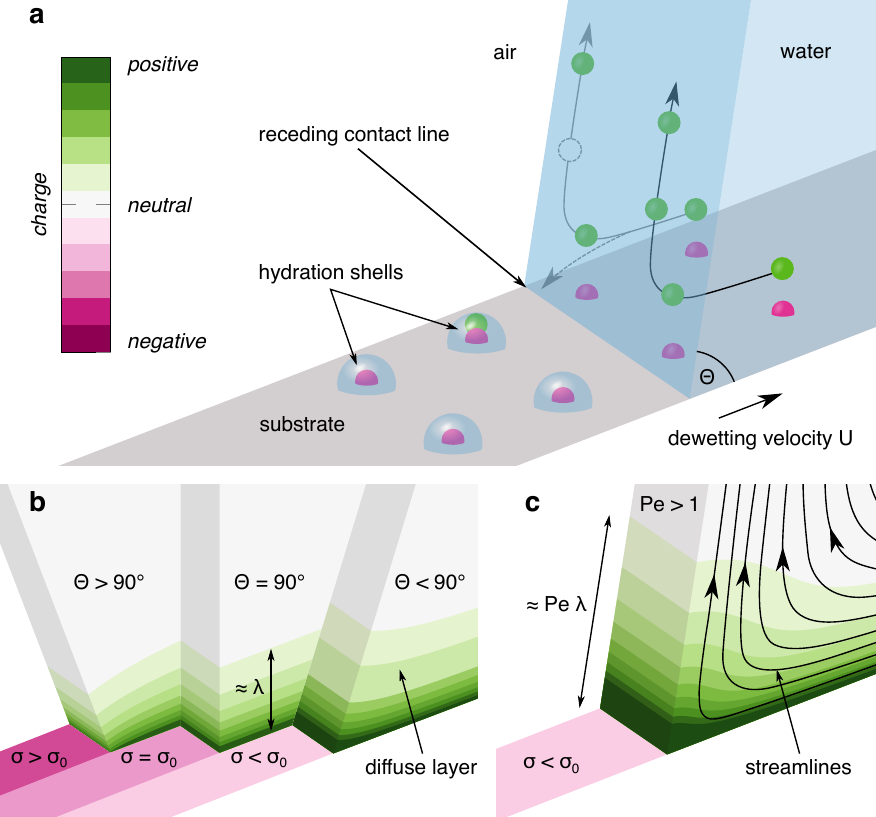}
  \caption{\textbf{Schematic representation of the charge separation.} All panels show the receding liquid wedge close to the three-phase contact line. Negative and positive and charges are represented by green and pink color, respectively. 
\textbf{a}, Essential charge separation mechanism. Bound surface charges are transferred from the wetted to the dewetted region as the three-phase contact line recedes. These charges are surrounded by hydration shells, some of which contain a counter charge and thus neutralize.
\textbf{b}, Contact angle effect. The EDL structure close to the contact line is sketched for contact angles $\theta$ smaller than, equal to, and greater than $\pi/2=90$° for low dewetting velocities. The isopotential lines in the diffuse layer (characterized by the Debye length $\lambda$) are warped due to the presence of the gas-liquid interface, which yields increasing surface charge densities $\sigma$ for increasing contact angles (neglecting effects on the atomistic scale).
\textbf{c}, Flow effect for high dewetting velocities and thus Péclet numbers greater than one. Advective transport along the streamlines that are parallel to the gas-liquid interface expands the diffuse layer.} \label{Fig:1}
\end{figure}

\paragraph{Factors influencing charge separation --}
Prerequisite for charge separation is the electric double layer (EDL) that forms on an initially uncharged solid surface when brought into contact with a liquid electrolyte such as water. The EDL is composed of 
bound surface charges and a diffuse layer of counter charges in the liquid with a thickness of $\approx1-\SI{300}{nm}$, called Debye length $\lambda$. Charge separation encompasses two essential steps. 
First, the surface acquires its bound net charge, screened by the diffuse layer. Second, the macroscopically electroneutral EDL separates at the receding contact line 
and a net charge remains on the dewetted surface, while a counter charge accumulates in the liquid (Fig. \ref{Fig:1}a).
Surface chemistry, contact angle, 
and fluid flow determine the EDL structure at the contact line and thus influence charge separation.

To understand charge separation in detail, we consider an initially wetted part of the surface to explore how it charges, how the EDL changes as the contact line approaches, and how much charge remains after dewetting. 

\paragraph*{Surface chemistry --} 
Several processes can lead to a charged surface \cite{Jacobs.1996}. 
For a number of surfaces (like \ce{SiO_2}), they can schematically be described by the reactions (see Methods) \cite{Sosa.2022} 

\begin{align}
    \ce{R-OH &<->[\textit{K}_{A}] R-O^- + H^+},\label{eq:deprotonation}\\ 
    \ce{H^+ + R-OH &<->[\textit{K}_{B}] R-OH_2^+},\label{eq:protonation}
\end{align} 
with the active surface sites $\ce{R-OH}$.
The surface charge will be screened by charges inside the liquid 
and can be quantified by the zeta potential $\zeta$, which is the electrostatic potential drop across the diffuse layer in equilibrium far from the contact line [see Supplementary Information (SI) \S2.1]. The surface chemistry yields a pH-dependent surface charge, specifically a point of zero charge (pzc) where the charge density vanishes. The constants $K_\mathrm{A}$ and $K_\mathrm{B}$ 
determine $\zeta$ and the pzc. 
When the local ion concentrations in the diffuse layer is disturbed, the equilibrium can shift and the potential drop can deviate from $\zeta$. We denote it $\phi$.
For $\mathrm{pH>pzc}$, the surface charge density $\sigma$ and potential $\phi$ are negative and governed by deprotonation (equation \ref{eq:deprotonation}) 
\cite{vanderWouden.2006}. 
Deprotonated surface sites
are in equilibrium with protons in the liquid, whose concentration depends on the potential, which is usually expressed relative to the thermal potential $\phi_\mathrm{T}=kT/e\simeq\SI{25}{mV}$ ($k$: Boltzmann constant, $T$: temperature, $e$: elementary charge). With the Debye-H\"uckel approximation $\phi/\phi_\mathrm{T}< 1$, the potential drop across the diffuse layer due to deprotonation (equation \ref{eq:deprotonation}) follows (see SI \S2.1)

\begin{equation}
    \phi=-\frac{C\phi_\mathrm{T}}{1+K^{-1}\left(1 -\phi/\phi_\mathrm{T}\right)}, \label{Eq:psi_implicit}
\end{equation}
which, far from the contact line is the zeta potential $\zeta$. 
The equilibrium constant and zeta potential are linked by $K=(\zeta/\phi_\mathrm{T}-1)/(C\phi_\mathrm{T}/\zeta+1)$ (see SI \S2.1).
Here, $C=e\Gamma\lambda/(\varepsilon \phi_\mathrm{T})$ ($\varepsilon$: liquid permittivity) 
is a non-dimensional measure for the active site density on the surface $\Gamma$. 
The surface charge far from the contact line,
$\sigma_0=\varepsilon \zeta /\lambda$, 
quantifies the charge that can potentially be separated in the dewetting process. Closer to the contact line, the surface charge is influenced by the contact angle, flow, and hydration effects (Fig. \ref{Fig:1}). We find that these effects are largely independent of each other and can be understood separately, starting with the contact angle.
 
\paragraph{Contact angle effects --}
We conceive the liquid shape at the contact line as a wedge \cite{Huh.1971}. 
Fig. \ref{Fig:1} (b) shows the EDL structure 
in the liquid close to the contact line, obtained from simulations (see Methods and SI \S1). For a contact angle $\theta=\pi/2$, the contours indicating constant values of the electrostatic potential are planar and the EDL structure is the same as far away from the contact line. 
For contact angles $\neq \pi/2$ however, the contours are significantly warped by the liquid-gas interface, where the normal electric field is negligible because of the jump condition for the electric field at the interface between two dielectric media and the high relative permittivity of water $\varepsilon_w\gg1$. 

In liquids, the diffuse layer screens the surface charge and establishes electroneutrality. Far from the contact line, the counter charge to the local surface charge distributes one-dimensionally in wall-normal direction. Yet, close to the contact line the counter charge distribution is two-dimensional. For $\theta>\pi/2$, the counter charge 
distributes over a larger angular domain, and for $\theta<\pi/2$ over a smaller one. Because larger (smaller) angular domains can accomodate more (less) counter charge, the local surface charge close to the contact line increases for $\theta>\pi/2$ and decreases for $\theta<\pi/2$, even at the same wall potential. The influence of the contact angle on the surface charge density was analyzed quantitatively by \cite{Dorr.2012}, who derived the ratio of nondimensional wall potential and surface charge density, $g(\theta)=\pi/(2\theta)$, for angles around $\pi/2$.
Assuming an approximately constant wall potential $\phi=\zeta$
(see SI \S2.3), 
the surface charge density at the contact line on the liquid side becomes 
\begin{equation}
    \sigma_\mathrm{CL}(\theta)=\frac{\varepsilon \zeta}{\lambda g(\theta)}. \label{Eq:sigma-theta}
\end{equation}

This purely geometrical effect is present even at negligible velocities. 

\paragraph{Flow effects --}
Since liquid adheres to the solid surface, contact line movement induces a flow in the wedge-shaped liquid domain \cite{Huh.1971}. 
The streamlines follow the solid-liquid and the liquid-gas interfaces, switching directions close to the contact line (Fig. \ref{Fig:1}c). Consequently, far from the contact line the flow is wall-parallel and 
does not significantly affect the local EDL structure. However, close to the contact line it changes to wall-normal direction. The normal flow affects ions in the EDL and modifies the diffuse layer. For the following arguments, we consider a frame-of-reference co-moving with the contact line.

To characterize this advective influence, we introduce the Péclet number 
that measures the relative importance of advective over diffusive transport. 
The Péclet number is defined as $Pe=U\lambda/D$, with the Debye length $\lambda$ as the only local length scale, ion diffusivity $D\approx\SI{1e-9}{m^2/s}$ and dewetting velocity $U$. It is $\approx1$ for $U=\SI{1}{cm/s}$ and $\lambda=\SI{100}{nm}$ and indicates the influence of dewetting velocity. Because of mass conservation, the velocity directly along the liquid-gas interface is essentially $U$ (Fig. \ref{Fig:1}c) and points wall-normal  for contact angles $\theta\simeq\pi/2$. Under the Debye-H\"uckel approximation, the ion distribution in the diffuse layer is governed by an advection-diffusion equation that can be reduced to one dimension for dominant wall-normal flow \cite{Ratschow.2022}. Analytically solving said equation (see SI \S2.1) reveals
an exponential decay of space charge and electrostatic potential over an effective length $\lambda_\mathrm{eff}$,
\begin{equation}
    \lambda_\mathrm{eff}=\frac{2}{\sqrt{Pe^2+4}-Pe}\lambda. \label{eq:lambdaeff}
\end{equation}
Evidently, the flow expands the diffuse layer close to the contact line (Fig. \ref{Fig:1}c). We observe $\lambda_\mathrm{eff}(Pe\ll1)\simeq\lambda$ and $\lambda_\mathrm{eff}(Pe\gg1)\simeq Pe\lambda$ and therefore predict two distinct regimes as a function of velocity that we will discuss later. 
 
\paragraph{Model for the surface charge density at the contact line --} We have established that the EDL structure depends on the contact angle and that the flow can expand the diffuse layer. 
Building on this understanding, we can formulate a model for the electrostatic potential and surface charge density at the contact line. We introduce the advection effect into the surface chemistry equation \ref{Eq:psi_implicit} and equation \ref{Eq:sigma-theta} for the surface charge density by using the effective Debye length (equation \ref{eq:lambdaeff}) and obtain the full analytical model  (see SI \S2.1)
\begin{eqnarray}
\begin{aligned}
&\frac{\phi_\mathrm{CL}(Pe)}{\phi_\mathrm{T}}=\frac{1}{2}\left(K+1\right)-\sqrt{ \frac{1}{4}\left(K+1\right)^2+KC\lambda_\mathrm{eff}/\lambda},\\
    &\sigma_\mathrm{CL}(\theta,Pe)=\frac{\varepsilon\phi_\mathrm{CL}  }{\lambda_\mathrm{eff} g(\theta)}. \label{Eq:fullmodel}
\end{aligned}
\end{eqnarray}
The surface charge in the liquid far from the contact line is $\sigma_0=\sigma_\mathrm{CL}(\theta=\pi/2,Pe=0)$. We deduce that when the contact line approaches, the surface charge gradually changes from $\sigma_0$ to $\sigma_\mathrm{CL}$. It suggests itself to assume that the surface charge $\sigma_\mathrm{CL}$ is dewetted at the receding contact line and remains on the surface. 
However, this neglects the effects occurring on the atomistic scale.

\paragraph{Effects on the atomistic scale --}
Assuming local thermodynamic equilibrium at the contact line, the probability that a dry ion remains on the dewetted surface instead of staying in the liquid is linked to its potential energy difference $\Delta U$ by the Boltzmann factor $\mathrm{exp}[-\Delta U/(kT)]$. Comparing the difference in Born solvation energy of a surface-bound ion in water vs. air reveals $\Delta U\gg kT$ \cite{Stetten.2019}, from which we infer that the probability is virtually zero. Yet, charge separation is observed in experiments 
\cite{Lin.2014,Sun.2015,Boamah.2019,Xu.2020,Wu.2020,He.2019,Shahzad.2018,Stetten.2019,Sun.2019,Lin.2020,Choi.2013,Nauruzbayeva.2020,Poli.2020,Armiento.2022,Li.2022}.
We thus speculate that surface bound charges leaving the liquid retain a  thin shell of water molecules, a hydration shell (Fig. \ref{Fig:1}a). These molecules could be condensed ambient humidity or originate from the dewetting liquid \cite{Stetten.2019} which contains a high counter-ion concentration to the surface charge.
When some counter-ions re-emerge in the hydration shells on the otherwise dry surface, the apparent net surface charge is diminished and thus always $\leq\sigma_\mathrm{CL}$. 
This effect likely depends on the specific types of ions and the chemical composition of the solid surface, among others, and quantification would require complex molecular dynamics simulations.
Therefore, the theoretical model of equation \ref{Eq:fullmodel} describes the universal part of the charge transfer mechanism, which needs to be supplemented by a description of the specific processes occurring on the atomistic scale. Comparisons of solutions of both our analytical model and detailed numerical simulations with experiments indicate that the net fraction $\omega$ of $\sigma_\mathrm{CL}$ on the surface lies in the range $0.1-1$. Notably, $\omega$ appears to be independent of $\theta$ and $Pe$, so the universal part of the theoretical model and the specific description of atomistic processes are independent and our model should be able to predict trends observed in experiments.

\begin{figure}
  \includegraphics[width=0.99\columnwidth]{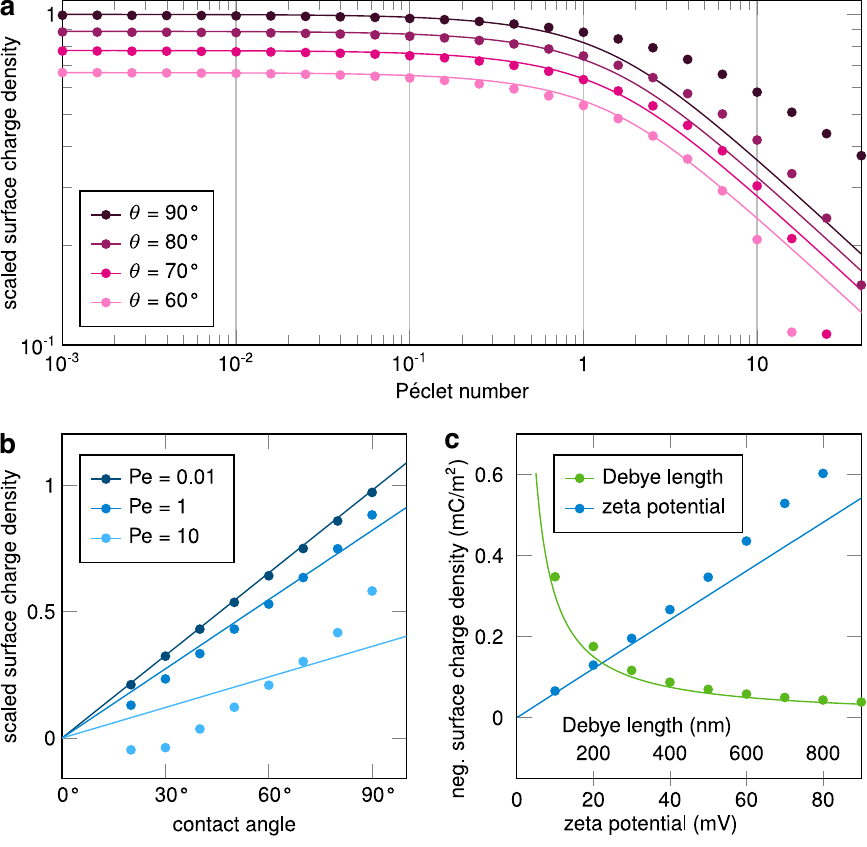}
  \caption{\textbf{Parametric dependencies of the surface charge density at the contact line on the liquid side} The graphs compare analytical (lines) with numerical results (symbols). All calculations use a Debye length of $\lambda=100~\si{\nano\meter}$, a zeta potential of $\zeta=-50~\si{\milli\volt}$ and a receding contact angle of $\theta=\ang{80}$, if not stated otherwise. The Péclet number $Pe$ is varied via the dewetting velocity $U$. With constant effects on the atomistic scale ($\omega=\mathrm{const.}$), the scaled surface charge density is a direct measure for charge separation. 
\textbf{a}, Scaled surface charge density at the contact line $\tilde{\sigma}$ as a function of the Péclet number at different contact angles $\theta$. 
\textbf{b}, Scaled surface charge density at the contact line $\tilde{\sigma}$ as a fuction of the receding angle at different Péclet numbers.
\textbf{c}, Negative value of the surface charge density at the contact line $-\sigma_{CL}$ as a function of the Debye length $\lambda$ (green) and  the zeta potential (blue) at $Pe=0.01$.
} \label{Fig:2}
\end{figure} 
 
\begin{figure}
  \includegraphics[width=0.99\columnwidth]{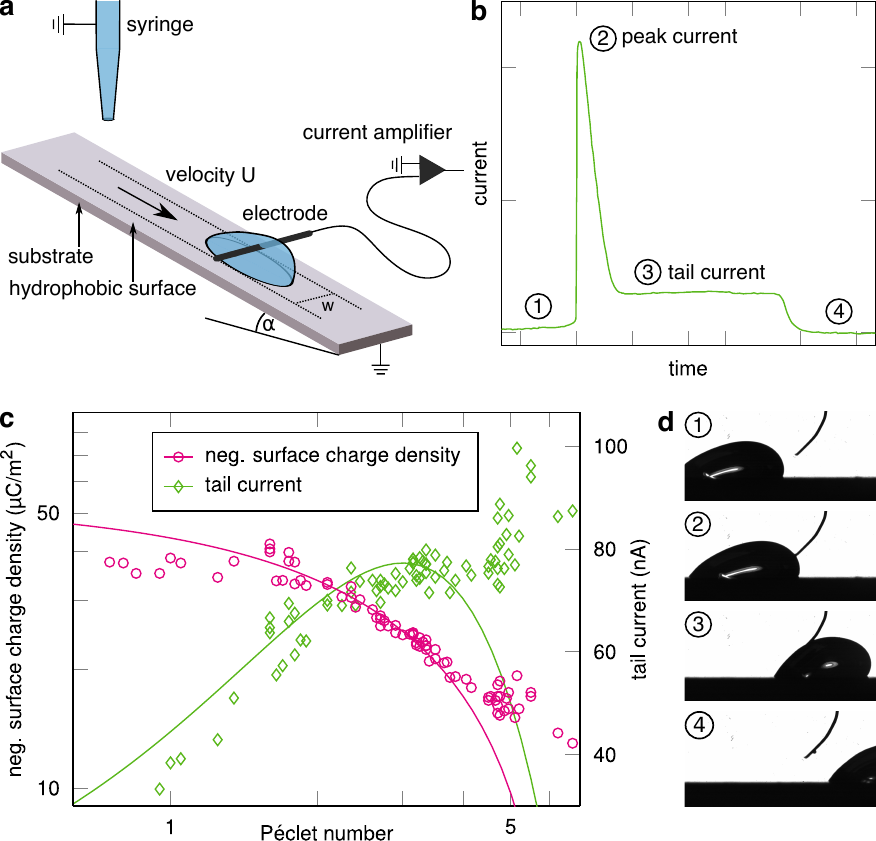}
  \caption{\textbf{Validation by means of experimental measured slide electrification.} 
\textbf{a}, Schematic of experimental setup. Drop slides down an inclined hydrophobic surface and accumulates a charge due to charge separation at the contact line. When touching the electrode, the drop gets discharged and the arising current is recorded.
\textbf{b}, Typical course of measured current for the contact time between drop and electrode with corresponding photographs in \textbf{d}. The peak current arises due to the discharge of the drop charge accumulated over the slide path before grounded by the electrode. The tail current measures the local deposited charge on the dewetted surface.
\textbf{c}, Comparison of analytical (line) and experimental results (dots) for different Péclet numbers $Pe$. The experimentally measured tail current $I_{tail}$ is drawn in green while the calculated negative value of the surface charge density left by the drop $-\sigma_\omega$ ($w=5~\si{\milli\metre}$, $\omega=0.37$) is plotted in pink.}\label{Fig:3}
\end{figure}

\paragraph{Predictions and implications --}
Charge separation is highest on hydrophobic surfaces (Fig. 2a-b). The higher the receding contact angle, the higher the scaled surface charge, which is the ratio of surface charge at the contact line and far from it and directly measures charge separation. In terms of the Péclet number we observe two distinct regimes (Fig. 2a). 
Up to $Pe \simeq 1$, charge separation is not influenced. Originally, we expected it to be a non-equilibrium process and suspected stronger charge separation when the flow drives the system further from equilibrium. However, counter-intuitively, from $Pe \simeq 1$ on, advective effects decrease charge separation with increasing velocity. This dependence is opposite to what is observed in flow electrification \cite{Gibson.1970,Zdanowski.2019} or solid-solid contact electrification \cite{Kaponig.2021} and caused by the extension of the diffuse layer, which reduces surface charge. 
Surfaces with higher zeta potentials experience higher surface charges after dewetting, while an increase in Debye length decreases charge separation (Fig. 2c). \\
Note that the presented model neglects electric fields in the substrate, corresponding to a grounded liquid (compare Fig. \ref{Fig:3}a). A more general model without this limitation is presented in the SI \S2.1.
\\

\paragraph{Numerical validation --}

The analytical model is based on a number of simplifying assumptions (see SI \S2.2).
To assess their applicability, we compare the analytical predictions to detailed numerical simulations (Methods and SI \S1). In Fig. \ref{Fig:2}a we see an excellent agreement of the two data sets for the scaled surface charge $\sigma_\mathrm{CL}/\sigma_0$ up to $Pe\simeq 1$. For higher $Pe$, still an agreement is found with respect to prediction of qualitative trends. 
In this $Pe$ regime, the potential drop at the contact line $\phi_\mathrm{CL}$ substantially increases compared to the zeta potential. Besides the increasingly inaccurate Debye-H\"uckel linearization, the surface chemistry is then influenced by proton adsorption, neglected in the analytical model. 
Even though the expression for the contact angle influence $g(\theta)$ is linearized around $\theta=$90° \cite{Dorr.2012}, it holds well for angles down to 20° (Fig. \ref{Fig:2}b), again with increasing deviations at higher $Pe$. For low contact angles and $Pe=10$ the proton concentration and thus the local pH can exceed the pzc, where proton adsorption dominates and the surface charge polarity flips, not captured by the analytical model.  In Fig. \ref{Fig:2}c we show that the model fairly accurately captures the dependence on zeta potential and Debye length, even beyond the limit of the Debye-H\"uckel approximation of $\zeta<\phi_\mathrm{T}\simeq\SI{25}{mV}$. 
Overall, the modelling assumptions appear justified as the deviations compared to simulations are smaller than $\omega$ (hydration effects) and the trends remain correct. 

\paragraph{Experimental validation --}
To validate the proposed theory of charge separation, we compare our model, equations \ref{Eq:fullmodel}, to experimental data. In the experiments, schematically shown in Fig. \ref{Fig:3}a and described under Methods, a 1mM $\mathrm{KNO_3}$ aqueous drop slides down an uncharged glass substrate with a hydrophobic coating. We optically measure the slide velocity and control it via the inclination angle $\alpha$. 
The drop slides along a tungsten wire electrode grounded through a femtoampere meter. When it first touches the electrode, the drop discharges (see Fig. 3b, peak current). While in contact, the drop is at ground potential but continuously deposits surface charge, measured by the tail current $I_\mathrm{tail}$. In Fig. \ref{Fig:3}c we compare the deposited surface charge density, $\sigma_\omega=I_\mathrm{tail}/(Uw)$ ($U$: drop velocity, $w$: drop width) to our model predictions, where we use the Cox-Voinov model (CVM) \cite{Voinov.1977,Cox.1986} to account for the dynamic changes of the receding contact angle. Hydration effects are represented by the fitting parameter $\omega$. With $\omega=0.37$ we obtain a fair agreement between theory and experiments up to $Pe \simeq 4$. Deviations for larger Péclet numbers are expected, since the CVM becomes inaccurate at non-negligible Reynolds numbers (see Methods). The decreasing charge separation at higher velocities is immediately apparent in the measurements.

\paragraph{Conclusions and outlook --}

To conclude, we propose that charge separation occurs by dewetting of the bound surface charge in the electric double layer while the diffuse counter charges remain in the liquid. The deposited charge increases with the zeta potential. Charge separation is strongest on hydrophobic substrates with high contact angles and -- contrary to other contact electrification mechanisms -- decreases with the dewetting velocity. Our analytical model quantitatively captures these parametric trends, in agreement with numerical simulations and experimental results. Based on the insights into the physics of charge separation reported in the present paper, the relevant parametric dependencies have been identified and can be probed in further experiments. We hypothesize that our results could open a fresh perspective on a number of findings that are difficult to explain without considering charge separation. For example, they could provide an additional explanation for contact angle hysteresis and changes in wetting properties after initial dewetting \cite{Mugele.2005} due to electrostatic forces between surface charge and liquid. Last but not least, by lifting some of the simplifying assumptions, our theory could be extended in different directions and could therefore form the nucleus of a class of models for charge separation by dewetting.

\begin{acknowledgments}
We wish to thank Maximilian T. Sch\"ur for helpful discussions on the simulations, and Xiaomei Li for providing the experimental photographs. 
This work was supported by the German Research Foundation (DFG) within the Collaborative Research Centre 1194 “Interaction of Transport and Wetting Processes”, Project- ID 265191195, subproject A02b (S. Hardt), the Department for Process and Plant Safety of Bayer AG, Leverkusen, Germany (A. D. Ratschow), and the European Research Council (ERC) under the European Union’s Horizon 2020 research and innovation program (grant agreement no. 883631) (P. Bista and H.-J. Butt).
\end{acknowledgments}
S.H. and H-J.B. proposed the work, A.D.R. developed the theoretical framework and the analytical model, L.S.B. carried out the simulations and proposed the measurement method, P.B. and S.A.L.W. conducted the experiments, A.D.R., L.S.B., S.H. and H-J.B. contributed to the interpretation of the results, A.D.R., L.S.B. and S.H. prepared the manuscript, and S.H. and H-J.B. supervised the work.

\bibliographystyle{apsrev4-1} 
\bibliography{main.bbl}  

\section{Methods}

\subsection{Experimental Setup}
\paragraph{Sample preparation --}
The hydrophobic samples consist of glass slides (25~x~70~x~1~mm) coated with trichloro(1H,1H,2H,2H-perfluorooctyl)silane (PFOTS) (Sigma-Aldrich 
Chemie GmbH) using chemical vapor deposition. Before coating, the substrates are cleaned with 
acetone and ethanol, and treated in an oxygen plasma cleaner (Diener 
Electronics Plasma surface, Femto BLS) for 10 minutes 
at 300~W to activate the surface. Then, the prepared slides and a 1~mL vial of PFOTS are placed into a vacuum desiccator, which is evacuated to a pressure 
of 100~mbar. The resulting hydrophobic surfaces have 
advancing and receding contact angles of 107~$\pm$~2° and 89~$\pm$~3°, 
respectively.\\ 

\paragraph{Setup and procedure --}
All experiments are performed in an inverted nitrogen atmosphere under ambient conditions (temperature: $21\pm1$~°C, humidity: $50\pm2~\%$). For each experiment, we place the hydrophobic sample on a grounded metal 
plate inside the humidity chamber, neutralize the 
surface by an ionizing air blower (Mini Zero Volt Ionizer 2, ESD) switched on for 5 
minutes, and wait some minutes until the ions in the air are
dissipated and have reached an equilibrium concentration. Then, a peristaltic pump (Minipuls 3, Gilson) produces a drop with a volume of $45~\mu$L consisting of 1 mM \ce{KNO3} solution (pH $\approx$ 5.5). After falling $0.5\pm0.2$~cm from the grounded syringe needle, the drop slides a distance of $5.0\pm0.2$~cm where it gets discharged by a measuring electrode. A transimpedance current amplifier (rise time: 
0.8~ms) (DDPCA-300, Femto) and a light barrier (laser diode CPS186, 
Thorlabs) placed $1.0\pm0.2$~cm in front of this electrode, allow recording the arising electrode current by a data acquisition board (USB-6366 x-Series, NI). While the peak of the signal is accounted to the discharge of the previously accumulated drop charge, the mean tail current gives the charge separation rate of grounded drops, see \ref{Fig:3}b. The velocity $U$ is calculated by means of the time elapsed between the laser trigger and the start of the discharge current at the electrode. Since the humidity chamber is installed on a tilting stage, the velocity is adjustable via the inclination angle. 

\paragraph{Evaluation --}
We translate the experimentally measured velocities $U$ to corresponding Péclet numbers $Pe=U\lambda/D$. The addition of $c_0=\SI{1}{mM}$ \ce{KNO3} allows us to determine the Debye length by $\lambda=\sqrt{\epsilon RT/(2z^2c_0F^2)}=\SI{9.7}{nm}$, with an ion valence of $z=1$ and the gas and Faraday constants $R$ and $F$. The salt diffusivity of \ce{KNO3} at low concentrations is approximately $D=\SI{2e-9}{m^2/s}$ \cite{Daniel.1991}. The experimental velocity range of $15-\SI{134}{cm/s}$ thus yields Péclet numbers of $0.75-6.7$. The zeta potential of the specific substrate used, $\zeta_0=\SI{-35}{mV}$, was measured by \cite{Vogel.2022}. It is rescaled according to $\zeta=\zeta_0\lambda/\lambda_0$ \cite{Kirby.2004}, where $\lambda_0=\SI{173}{nm}$ is the Debye length of DI water with a \ce{CO2} concentration in equilibrium with atmospheric \ce{CO2}, $\mathrm{pH}=5.5$ as used by \cite{Vogel.2022}. The active site density on the substrate, $\Gamma=\SI{5}{nm^{-2}}$, was reported by \cite{Vogel.2022,Christl.1999,Hal.1996}.

\subsection{Numerical Simulations}

\paragraph{Computational domain --}
The numerical simulations are performed using Comsol Multiphysics, version 6.1, which is based on the finite-element method. Assuming that the charge accumulation is caused by the separation of the diffuse layer from the immobilized charges at the surface while dewetted, the computational domain is reduced to a section close to the receding contact line. Thus, the computational domain contains the dewetting liquid idealized as a two-dimensional wedge shaped geometry.
This wedge has an opening angle of $\theta$ and a radius of $L=100\lambda_\mathrm{DI-water}$. To eliminate finite-size effects, we expand the computational domain by a predomain with length $L_\mathrm{pre}=c_\mathrm{pre}L$. The liquid surface is assumed to be perfectly clean, flat and smooth. We consider a frame-of-reference co-moving with the contact line.\\  

\paragraph{Governing equations and boundary conditions --}The mathematical model consists of the Stokes and Poisson-Nernst-Planck (PNP) equations for an incompressible, Newtonian liquid, where the electric body force is neglected as its influence is below the numerical accuracy.\\

At the liquid-gas interface, we prevent flow and species flux across the interface.  Furthermore, we account for the low viscosity and relative permittivity of gas in comparison to aqueous solutions and apply zero tangential shear stresses and a zero normal electric field.\\
Along the wall, the moving contact line is modeled by a tangential wall velocity $\mathbf{u}_w$ which equals the drop velocity $U$ in experiments. To avoid singularities at the contact line, the Navier slip boundary condition with a slip length of $l_\mathrm{s}$ is applied to the tangential velocity component \cite{Dussan.1974}.
Normal flow across the solid-liquid interface is inhibited. However, the flux of individual species
is non-zero since it is governed by adsorption and dissociation processes at the wall and thus, the surface chemistry must be taken into account. Specifically, a charge regulation model including two reaction equations is employed, which is described in more detail in the paragraph \textit{surface chemistry}. 
Additionally, when considering a local surface element, the surface charge moves along the wall due to the co-moving frame-of-reference. Consequently, we complement the boundary conditions by a surface charge conservation equation.
It ensures that the change in surface charge along the solid-liquid interface is balanced by the charge transferred from the liquid. 
At the arc-shaped boundary, which represents the boundary to the bulk liquid, an isobaric boundary condition with a reference pressure of $p_0=0$ is applied. Additionally, at this boundary we approximate the potential and species concentrations by the solution of the Poisson-Boltzmann equation for an infinite flat plate. All governing equations and the corresponding boundary conditions are given in the SI \S 1.\\

The discretization error is limited to 0.5\% based on a grid convergence study, and the simulations were checked for finite-size effects. Moreover, the effects of the disregarded inertial terms and electric volume force in the Stokes equation were quantified. The corresponding changes in the results are much smaller than the numerical accuracy, which is why these effects can be neglected. \\

\paragraph{Surface chemistry --} \label{sec:suface_chemistry}
The surface chemistry is modeled by a charge regulation model using two reaction equations with equilibrium constants determined from the known point of zero charge and the zeta potential. The model considers surface charging based on ion adsorption and dissociation \cite{Sosa.2022}. We assume a fixed number of occupyable, indistinguishable, amphoretic surface groups, which can each bind one molecule. Furthermore, we assume a symmetrical electrolyte.
Exemplary reaction equations read
\begin{align}
    \ce{R-OH &<->[\textit{K}_{A}] R-O^- + H^+}, \label{eq:methodsdeprotonation}  \\ 
    \ce{H^+ + R-OH  &<->[\textit{K}_{B}] R-OH_2^+}, \label{eq:methodsprotonation}
\end{align} 

where hydroxide surface groups react with DI water. The symbol R denotes the active surface groups. We introduce the occupancy rates for negatively and positively charged sites $\alpha$ and $\beta$,  defined as
\begin{align}
		\alpha:=\frac{[\mathrm{R-O^-}]}{\Gamma}\quad\text{and}\quad
		\beta:=\frac{[\mathrm{R-OH^+_2}]}{\Gamma},
\end{align}
where the brackets indicate surface concentration and $\Gamma$ is the concentration of active surface groups. 
Further, we assume the surface reaction to be always in equilibrium. This is justified, since the time scale of the dissociation process is much shorter than the typical contact time of surface and drop. Even a very high drop velocity of $1$ $\mathrm{ms}^{-1}$ and a drop diameter of $5~\mathrm{mm}$ yield a time scale of $\tau_\mathrm{contact}=5~\mathrm{ms}$, while typical timescales of dissociation processes are of the order of $\tau_\mathrm{dis}=10^{-3}~\mathrm{ms}$ \cite{Jacobs.1996}. 

The law of mass action for the two reactions with equilibrium constants $K_\mathrm{A}$ and $K_\mathrm{B}$ reads
\begin{align}
	\begin{split}
	K_\mathrm{A}=\frac{\alpha c_+}{(1-\alpha-\beta)}\quad\text{and}\quad
	K_\mathrm{B}=\frac{\beta}{(1-\alpha-\beta)c_+}.
	\end{split}\label{eq:20_GGW_const_red}
\end{align}
where $c_+$ denotes the proton concentration. These equations \ref{eq:20_GGW_const_red} can be solved for $\alpha$ and $\beta$. Since the surface charge density $\sigma_s$ is the difference in charge density between the oppositely charged surface groups, it can be expressed in terms of the equilibrium constants and the proton concentration
\begin{align}
	\sigma_s=\Gamma z e (\alpha-\beta)=\Gamma z e \frac{K_\mathrm{B}c_+^2-K_\mathrm{A}}{K_\mathrm{B}c_+^2+c_++K_\mathrm{A}} \label{eq:20_surface_charge}
\end{align}
As pointed out in the main text, 
the equilibrium constants $K_\mathrm{A}$ and $K_\mathrm{B}$ determine the point of zero charge and the zeta potential. When (i) the point of zero charge (pzc) and (ii) the zeta potential for a given pH-value are known, the equilibrium constants can be recovered from these parameters. 
The pzc defines the pH value and thus the bulk ion concentration $c_{pzc}$ at which the surface chemistry adjusts to a zero surface charge. 
For the second condition, the corresponding surface charge density needs to be computed from the zeta potential. The charge density is related to the normal electric fields via $\sigma_s/\varepsilon_0=\mathbf{n}\cdot(\varepsilon_\mathrm{l}\mathbf{E}_\mathrm{l}-\varepsilon_{s}\mathbf{E}_\mathrm{s})$. When the relative permittivity of the solid $\varepsilon_\mathrm{s}$ is much smaller than that of the liquid $\varepsilon_\mathrm{l}$, the electric field in the solid $\varepsilon_\mathrm{s}\mathbf{E}_\mathrm{s}$ is negligible and the surface charge density is directly proportional to the electric field $\mathbf{n}\cdot \mathbf{E}_\mathrm{l}$ that can be determined from the analytical Gouy-Chapman potential distribution. The proton concentration at the wall corresponding to the bulk concentration $c_0$ is given by the Boltzmann factor $c_\mathrm{zeta}=c_0\exp(-\zeta/\phi_T)$.
The surface charge density together with the corresponding wall concentration in equation \ref{eq:20_surface_charge} for the situations (i) and (ii) recovers the equilibrium constants as 
\begin{align}
	    K_\mathrm{A}=K_\mathrm{B}c_{pzc}^2 \quad\text{and}\quad\\
	    K_\mathrm{B}=-\frac{c_\mathrm{zeta}}{\Gamma \lambda k_2(c_\mathrm{pzc}^2-c_\mathrm{zeta}^2)+(c_\mathrm{pzc}^2+c_\mathrm{zeta}^2)}, \label{eq:K_B}
\end{align}
with $k_{2}=ez/(4\varepsilon \phi_T)(\tanh^{-1}(\tilde{\zeta}/4)-\tanh(\tilde{\zeta}/4))$ and the dimensionless zeta potential $\tilde{\zeta}=\zeta/\phi_T$. 
It should be pointed out that the surface chemistry model and the corresponding boundary conditions for the numerical simulations, equations \ref{eq:20_GGW_const_red} - \ref{eq:K_B}, work equally well for ion adsorption and other surface groups.
It is intuitive that the surface charging mechanism itself is unimportant for the charge balance as long as we can assume the reaction being in equilibrium.

\subsection{Analytical calculations: Dynamic contact angles}
The analytical model captures the parametric dependencies on the receding contact angle as well as on the dewetting velocity measured by the Péclet number and treats the two effects independently. In general, dynamic receding contact angles are velocity dependent. To account for this dependency when comparing the model to experimental results (Fig. 3c), we use the Cox-Voinov model. It relates the dynamic receding contact angle $\theta$ to Young's contact angle $\theta_Y$ via $\theta^3=\theta_Y^3+9Ca\mathrm{ln}(l_M/l_\mu)$, with the capillary number $Ca=\eta U/\gamma$ ($\eta$: liquid viscosity, $\gamma$: liquid surface tension). We already measured and reported the Young's contact angle $\theta_Y=100$° for the PFOTS substrate used in \cite{Li.2022}. The expression $l_M/l_\mu$ is the ratio of the macroscopic and microscopic length scales, where the microscopic length scale $l_\mu$ is related to the size of molecules $\approx\SI{1}{nm}$. \cite{Voinov.1977,Cox.1986} As per \cite{Cox.1986}, the ratio is $l_M/l_\mu=10^{4}$. The model is only valid for low Reynolds numbers $<1$, expressed as $Re=Ul_M\rho/\eta$ ($\eta/\rho$: kinematic liquid viscosity). For our experimental conditions of $\lambda \approx \SI{10}{nm}$, the Reynolds number is approximately twice the Péclet number and thus the Cox-Voinov model loses validity for high $Re\simeq 2Pe$, as discussed in the main text.
 
\clearpage


\section*{Supplementary material}

# HOW CHARGES SEPARATE AT MOVING CONTACT LINES

## -Supplementary Information-

Aaron D. Ratschow,[1] Lisa S. Bauer,[1] Pravash Bista,[2] Stefan A. L. Weber,[2,3] Hans-Jürgen Butt,[2] and Steffen Hardt[1,*]

[1]*Institute for Nano- and Microfluidics, TU Darmstadt,*
*Alarich-Weiss-Straße 10, D-64237 Darmstadt, Germany*
[2]*Max Planck Institute for Polymer Research,*
*Ackermannweg 10, 55128 Mainz, Germany*
[3]*Department of Physics, Johannes Gutenberg University,*
*Staudingerweg 10, 55128 Mainz, Germany*
(Dated: May 3, 2023)

The Supplementary Information consists of two main sections. Section §1 reports additional information about the numerical model while section §2 contains further details regarding the derivation of the analytical model and the assumptions used.

## §1 NUMERICAL SIMULATION

This part completes the information on the numerical simulations, starting with the computational domain, the governing equations and boundary conditions in §1.1. We define all parameters used in the simulations in §1.2. The grid convergence study and the finite-size study are described in §1.3. The final section §1.4 takes a closer look at the sensitivity of the numerical results to variations in the parameters used for the boundary conditions.

### §1.1 Numerical Model

All simulations were performed with Comsol Multiphysics, version 6.1, which is based on the finite-element method. Fig. S4 shows the computational domain and the grid structure. The computational domain is discretized with a structured grid consisting of quadrilateral elements, which are refined towards the interfaces and the contact line based on geometric series. Additionally, a finely resolved boundary layer of cells with thickness $\lambda$ is defined at the solid-liquid interface. The velocity is discretized with quadratic and all other variables with linear shape functions.

The governing equations as well as the boundary conditions are shown in table S1, where $p$ is the pressure, $\mathbf{u}$ the velocity field, $\Phi$ the electric potential, $\mathbf{J}_i$ the species flux, $c_i$ the species concentration, $D_i$ the diffusivity and $z$ the species' valence. The subscripts $i = +$ and $i = -$ represent cations and anions, respectively. Additionally, $\varepsilon = \varepsilon_0 \varepsilon_l$ refers to the permittivity, $T$ to the temperature, $R$ to the gas constant, $e$ to the elementary charge and $F$ to the Faraday constant. The PNP equations are coupled by the space charge density $\rho_v = F \sum z c_i$ and the electric field is given by $-\nabla \Phi = \mathbf{E}$. The surface divergence operator is $\nabla_s = \nabla - \mathbf{n} \cdot (\mathbf{n} \cdot \nabla)$. Additional quantities and parameters are the shear stress $\tau = \eta[\nabla \mathbf{u} + (\nabla \mathbf{u})^{\mathrm{T}}]$, the wall velocity $\mathbf{u}_w$ and the drop velocity $U$. The slip length is denoted as $l_s$, the active surface group density is $\Gamma$ and the Debye length $\lambda$. The reaction constants $K_1$ and $K_2$ are fitted to the point of zero charge and the zeta potential. Their calculation is found in the main text under Methods.

### §1.2 Parameters

The parameters used are given in table S2. Since we restricted our calculations to symmetric binary electrolyte solutions, the bulk concentration is related to the Debye length $\lambda$

$$c_{i,0} = \frac{\varepsilon RT}{2(zF\lambda)^2} \quad . \tag{S1}$$

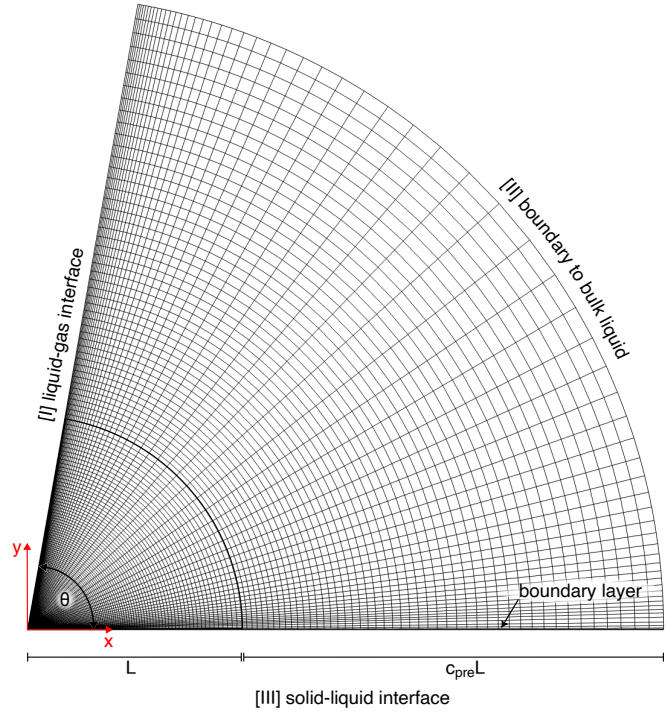

FIG. S1. Computational domain and schematic grid structure

TABLE S1. Governing equations and boundary conditions for the numerical simulations. Here, $\mathbf{n}$ and $\mathbf{t}$ denote outward normal and tangential vectors in a right-handed system and the coordinate $y$ is pointed normal to the solid-liquid interface [III], where it is zero.

| | | Stokes | Poisson-Nernst-Planck |
|---|---|---|---|
| **Governing equations** | | $0 = -\nabla p + \eta \nabla^2 \mathbf{u}$ | $\nabla^2 \psi = -\varepsilon^{-1} \rho_v$ |
| | | $\nabla \cdot \mathbf{u} = 0$ | $\nabla \cdot \mathbf{J}_i = 0$ |
| | | | $\mathbf{J}_i = c_i \mathbf{u} - D_i \nabla c_i - z D_i (RT)^{-1} F c_i \nabla \psi$ |
| **[I]** | **liquid-gas interface** | $\mathbf{u} \cdot \mathbf{n} = 0$ | $\mathbf{E} \cdot \mathbf{n} = 0$ |
| | | $\varphi = \theta$ | $\boldsymbol{\tau} \cdot \mathbf{t} = 0$ | $\mathbf{J}_i \cdot \mathbf{n} = 0$ |
| **[II]** | **boundary to bulk liquid** | $\boldsymbol{\tau} \cdot \mathbf{n} = 0$ | $\Phi = 2z^{-1}\phi_T \ln\left[\frac{1+\exp(-\lambda^{-1}y)\tanh(\tilde{\zeta}/4)}{1-\exp(-\lambda^{-1}y)\tanh(\tilde{\zeta}/4)}\right]$ with $\tilde{\zeta} = \frac{z\zeta}{\phi_T}$ |
| | | $r = (1 + c_{\text{pre}})L$ | $\boldsymbol{\tau} \cdot \mathbf{t} = 0$ | $c_i = c_{i,0} \exp\left(\frac{z\Phi}{\phi_T}\right)$ |
| **[III]** | **solid-liquid interface** | $\mathbf{u} \cdot \mathbf{n} = 0$ | $\sigma_s = \Gamma z e \frac{K_2 c_+^2 - K_1}{K_2 c_+^2 + c_+ + K_1}$, $\mathbf{E} \cdot \mathbf{n} = -\varepsilon^{-1}\sigma_s$ |
| | | $\varphi = 0$ | $u_{\text{slip}} = (\mathbf{u}_w - \mathbf{u}) \cdot \mathbf{t}$ | $-U \nabla_s \sigma_s = F z \mathbf{n} \cdot \mathbf{J}_+$ |
| | | | $u_{\text{slip}} = l_s \nabla_s \cdot \mathbf{u}$, $\mathbf{u}_w = U\mathbf{t}$ | $\mathbf{n} \cdot \mathbf{J}_- = 0$ |

## §1.3 Verification

In this section we provide insight into the numerical errors associated with the finite size of the computational domain and with discretization. The simulations performed to quantify these errors were carried out for a high Péclet number and a small Debye length, as seen in table S3, since the simulations are most susceptible to errors under these conditions.

We performed a grid convergence study to ensure grid independence of the numerical solutions. By defining one

TABLE S2. Reference conditions, electrolyte and surface properties for all simulations

| Parameter | Value | Description | Reference |
|---|---|---|---|
| $T$ | 298.15 K | Temperature | |
| $\eta\,(T = 298.15\text{ K})$ | $8.93 \cdot 10^{-4}$ Pa s | Dynamic viscosity | Comsol library |
| $\rho\,(T = 298.15\text{ K})$ | 998.21 kg/m³ | Density | Comsol library |
| $\varepsilon_\mathrm{r}\,(T = 298.15\text{ K})$ | 78.3 | Relative permittivity | [S1] |
| $D_-$ | $9 \cdot 10^{-9}$ m² s⁻¹ | Diffusion coefficient of anions | [S2, p. 918] |
| $D_+$ | $9 \cdot 10^{-9}$ m² s⁻¹ | Diffusion coefficient of cations | [S2, p. 918] |
| $l_\mathrm{s}$ | 1 nm | Slip length | [S3, S4] |
| $\Gamma$ | 5 nm⁻² | Active site density | [S5, S6] |
| pzc | 3.55 | Point of zero charge | [S7] |
| $\lambda_{DI-water}$ | 0.96 µm | Debye length of DI-water | |
| $z$ | 1 | species' valence | |

additional grid node between all existing ones, the grid was systematically refined with refinement factors of powers of 2. As convergence indicator, the surface charge density at the contact line was compared to its value obtained by Richardson extrapolation, and a grid with a relative error $\leq 0.005$ was chosen. The relative error of the numerical solver was set to 0.001.

To analyze finite-size effects, the predomain length $L_\mathrm{pre}$ was systematically expanded. Because an infinite predomain length is not numerically feasible, we used $L_\mathrm{pre} = 64L$ as an estimate and calculated the error relative to this value. Note that the associated error is $< 0.5\%$ for all predomain sizes considered (Fig. S2b). Thus, a value of $c_\mathrm{pre} = 4$ was chosen for all simulations.

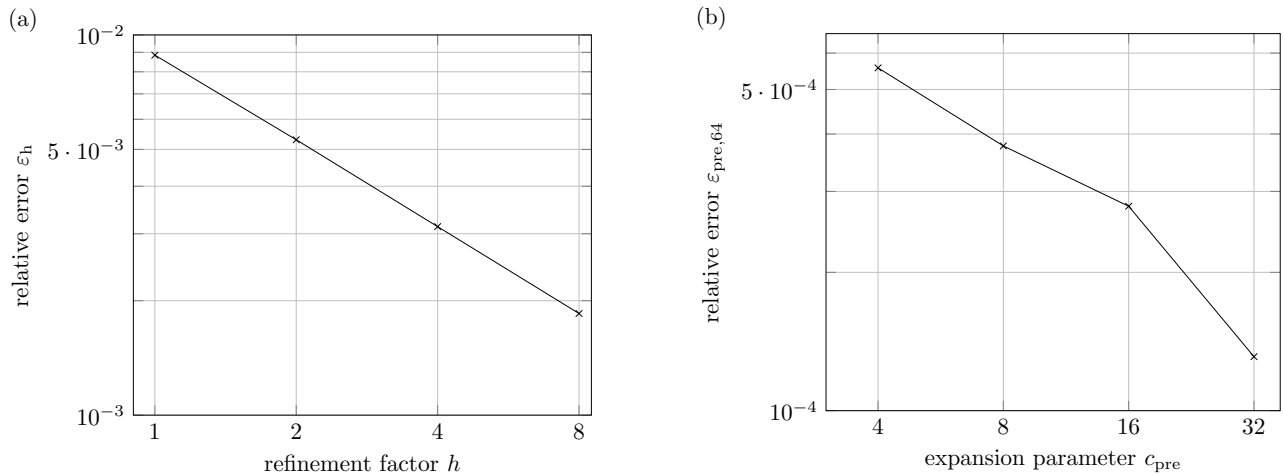

FIG. S2. a) Relative error $\delta_\mathrm{h}$ to the Richardson extrapolation for different grids with a refinement factor of $h$. The number of elements in each direction is $\propto h$ b) Relative error due to finite-size effects. By expanding the computational domain, the relative error $\delta_{\mathrm{pre},64}$ for each predomain size $c_\mathrm{pre}L$ is compared to the largest calculated domain with a predomain size of $64L$

TABLE S3. Simulation parameters used for the convergence and the finite-size study

| Parameter | Value | Description |
|---|---|---|
| $\zeta$ | $-50$ mV | $\zeta$-potential |
| $Pe$ | 40 | Péclet number |
| $\theta$ | $80°$ | Receding angle |
| $\lambda$ | 100 nm | Debye length |

## §1.4 Sensitivity to Parameters in Boundary Conditions

Often, the parameters characterizing a solid-liquid interface are difficult to obtain. Therefore, we would like to address the sensitivity of the surface charge density at the contact line with respect to variations in the surface characteristics, which include the active site density $\Gamma$ and the slip length $l_s$. We varied the active site density by considering the values $\Gamma \in \{0.1; 1; 10\}$ nm$^{-2}$. All yielded differences in the charge density at the contact line that are much smaller than the numerical accuracy. We conclude that uncertainties in this parameter are irrelevant for our simulations.

Contrary to this, variations of the slip length have a larger impact. Considering slip length variation of $l_s \in \{1; 3; 5; 10\}$ nm revealed changes in the charge density at the contact line of some percent in comparison to the smallest slip length of $l_s = 1$ nm, see fig. S3. This influence increases with increasing Péclet number. Nevertheless, we consider the variation induced by the slip length tolerable. This variation indicates how large the influence of atomistic effects in the close vicinity of the contact line may be. These effects cannot be fully captured using a continuum-mechanical description, which is why on this basis, a completely accurate description of charge transfer cannot be expected.

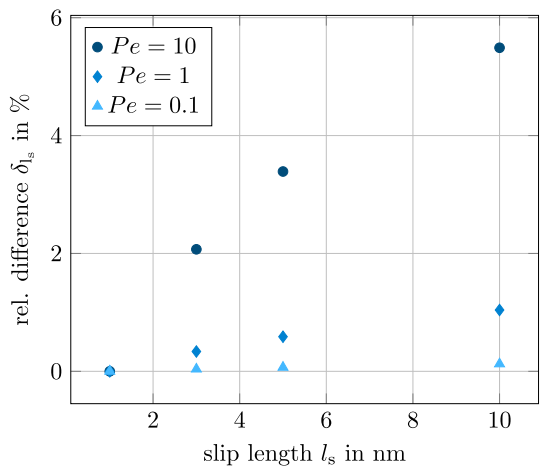

FIG. S3. Difference $\delta_{l_s}$ of the surface charge density at the contact line $\sigma_{CL}$ for different slip lengths $l_s$ at different Péclet numbers $Pe$ relative to the surface charge density for $l_s = 1$ nm

## §2 THEORY

### §2.1 Derivation

As described in the main text, our theory builds on the dewetting of bound surface charges within the EDL at the receding contact line as the fundamental mechanism of charge separation. Thus, the analytical model aims at quantifying this bound surface charge. To this end, we first discuss the surface chemistry, followed by contact-angle effects and flow effects. The model presented here also includes electric fields in the substrate due to a nonzero potential difference $V_l$ between the liquid and the substrate. It simplifies to the model presented in the main text when the liquid is at ground potential.

*Surface chemistry –* There are three main processes that can lead to a charged surface: (i) dissociation of surface groups, (ii) adsorption of ions to the surface via van-der-Waals forces or chemical bonds, and (iii) dissolution of the surface, which is negligible in the present context, as it happens on timescales $\approx 10^2 - 10^7$ s [S8]. For combinations of aqueous liquids and solids where specific adsorption of salt ions is negligible, like salty water on glass, the former two processes, (i) and (ii), can schematically be described by two reactions where the surface respectively gains one

positive or one negative net charge (see Methods, charge regulation model) [S9–S11]

$$\text{R--OH} \xleftrightarrow{K_\text{A}} \text{R--O}^- + \text{H}^+, \tag{S2}$$

$$\text{H}^+ + \text{R--OH} \xleftrightarrow{K_\text{B}} \text{R--OH}_2{}^+, \tag{S3}$$

with the active surface sites R−OH. For the analytical model we only consider deprotonation (equation S2), which governs the surface chemistry for pH $\gg$ pzc. We assume a number of active sites with a density $\Gamma$ on the surface, as well as chemical equilibrium, which is justified by the small timescale of the reaction $\approx 1 \times 10^{-6}\,\text{s}$ [S8]. The surface charge density $\sigma_0$ in chemical equilibrium is then given by the law of mass action as

$$\sigma_0 = -\frac{e\Gamma}{1 + n_{\text{H}^+}/K_\text{A}}. \tag{S4}$$

The local proton concentration in the liquid $n_{\text{H}^+}$ follows a Boltzmann distribution with the potential drop across the diffuse layer $\phi$ and the bulk proton concentration $n_0$

$$n_{\text{H}^+} = n_0 \exp\left(-\frac{\phi}{\phi_T}\right), \tag{S5}$$

from which we obtain

$$\sigma_0 = -\frac{e\Gamma}{1 + K^{-1}\exp(\phi/\phi_T)}, \tag{S6}$$

where $K = K_\text{A}/n_0$. This equation couples the surface charge density $\sigma_0$ and the potential drop across the diffuse layer $\phi$ through surface chemistry. Additionally, the jump condition for the electric field at the solid-liquid interface $\mathbf{n} \cdot (\varepsilon_l \mathbf{E}_l - \varepsilon_s \mathbf{E}_s) = \sigma_0/\varepsilon_0$ applies, with the normal vector $\mathbf{n}$, relative permittivity $\varepsilon$, and electric field $\mathbf{E}$ in the liquid $l$ and solid $s$. Under the Debye-Hückel approximation, $\phi < \phi_T$, the electric field in the liquid scales like $\mathbf{n} \cdot \mathbf{E}_l = \phi/\lambda$, with the Debye length for symmetric monovalent electrolytes $\lambda = \sqrt{\varepsilon_0 \varepsilon_l kT/(2e^2 n_0)}$. The electric field in the solid establishes between the liquid and a grounded sub-surface electrode (see Fig. 3a). Based on the picture of effective capacitors, the potential drop from the liquid to the subsurface electrode is due to a Stern layer capacitance and the dielectric capacitance of the substrate with thickness $d$. When the Stern layer capacitance is much larger than both the dielectric and the diffuse layer capacitance, as is the case for $d$ and $\lambda \geq 10\,\text{nm}$, it is negligible [S12]. Then, the electric field in the solid is $\mathbf{n} \cdot \mathbf{E}_s = -V_l/d$ and the interfacial boundary condition yields

$$\sigma_0 = \frac{\varepsilon_0 \varepsilon_l \phi}{\lambda} + \frac{\varepsilon_0 \varepsilon_s V_l}{d}. \tag{S7}$$

We define the nondimensional groups $C = e\Gamma\lambda/(\varepsilon_0 \varepsilon_l \phi_T)$ for the active site density and $V = \varepsilon_s \lambda V_l/(\varepsilon_l d\phi_T)$ for the potential inside the liquid and eliminate the surface charge density $\sigma_0$ from equations S6 and S7 to get an implicit equation for the potential drop across the diffuse layer

$$\phi = \frac{C\phi_T}{1 + K^{-1}\exp(-\phi/\phi_T)} - V\phi_T. \tag{S8}$$

This equation cannot be solved analytically for $\phi$. We thus linearize the exponential term in accordance with the Debye-Hückel approximation and arrive at

$$\phi = \frac{C\phi_T}{1 + K^{-1}(1 - \phi/\phi_T)} - V\phi_T, \tag{S9}$$

which is identical to equation 3 in the main text for $V_l = V = 0$. To retrieve the experimentally measureable value of the zeta potential for $V = 0$, the equilibrium constant has to be $K = (\zeta/\phi_T - 1)/(C\phi_T/\zeta + 1)$.

*Contact angle effect –* Dörr and Hardt [S13] derived an implicit analytical solution for the linearized Poisson-Boltzmann equation $\nabla^2\psi = 1/\lambda^2\psi$ in the vicinity of a contact line with contact angle $\theta$. They found that the equipotential lines warp close to the contact line due to the presence of the gas-liquid interface (Fig. 1b), where the boundary condition was approximated by a zero normal electric field due to the high permittivity of water compared to air. For contact angles around $\theta = \pi/2$ they found the generic relationship

$$\frac{\phi_{\text{CL}}}{\phi_0}\frac{\sigma_0}{\sigma_{\text{CL}}} = \frac{\pi}{2\theta} := g(\theta), \tag{S10}$$

where CL and 0 denote quantities at the contact line and far from it, respectively. In our simulations at low Péclet numbers, we only see minor changes in the surface potential towards the contact line, $\phi \simeq \zeta$ (section §2.3), and thus use the approximation $\sigma_0/\sigma_{\mathrm{CL}} = g(\theta)$. Consequently, the surface charge at the contact line becomes

$$\sigma_{\mathrm{CL}}(\theta) = \frac{\varepsilon_0 \varepsilon_l \zeta}{\lambda g(\theta)} + \frac{\varepsilon_0 \varepsilon_s V_l}{d}, \tag{S11}$$

which simplifies to equation 4 of the main text if $V_l = 0$. The expression for $g(\theta)$ could be replaced by a more exact expression, derived in [S14], which is valid for arbitrary contact angles. However, for the sake of simplicity, here we limit ourselves to the expression from equation S10.

*Flow effect* – Because liquid adheres to the solid surface, the movement of the contact line induces a flow in the receding liquid wedge [S15] (Fig. 1c). Viewed in the frame-of-reference co-moving with the contact line, the streamlines are parallel to the solid-liquid interface far from the contact line and re-orient to normal direction near the contact line for contact angles $\theta \simeq \pi/2$. Owing to mass conservation, the velocity along the solid-liquid interface and also the velocity along the gas-liquid are essentially the dewetting velocity $U$. Thus, near the contact line, wall-normal advective transport affects the charges in the diffuse layer.

The governing equations for ion transport are the Poisson-Nernst-Planck equations. For symmetric electrolytes, they can be transformed into the Poisson equation coupled to one equation each for the space charge density $\rho_v = e(n_+ - n_-)$ and the ionic conductivity $k = eD/\phi_T(n_+ + n_-)$. In the Debye-Hückel limit, the conductivity is constant. Then, the charge transport is governed by a single equation [S16],

$$\frac{\partial \rho_v}{\partial t} + \nabla \cdot (\rho_v \mathbf{u}) = D\nabla^2 \rho_v - \frac{k}{\varepsilon_0 \varepsilon_l}\rho_v, \tag{S12}$$

with the velocity $\mathbf{u}$. In its stationary form and evaluated in one dimension (wall-normal direction $y$), the equation reads

$$D\frac{\partial^2 \rho_v}{\partial y^2} - U\frac{\partial \rho_v}{\partial y} - \frac{k}{\varepsilon_0 \varepsilon_l}\rho_v = 0, \tag{S13}$$

where $\mathbf{n} \cdot \mathbf{u} = U$ is the wall-normal velocity. We analytically solve it alongside the one-dimensional Poisson equation $\partial^2 \psi/\partial y^2 = -\rho_v/(\varepsilon_0 \varepsilon_l)$, with the boundary conditions $\rho_v = \rho_{v,0}$ and $\psi = \psi_0$ at $y = 0$, as well as $\partial \rho_v/\partial y = \partial \psi/\partial y = 0$ at $y \to \infty$, and find the potential distribution

$$\psi = \psi_0 \exp\left(-y/\lambda_{\mathrm{eff}}\right), \tag{S14}$$

$$\lambda_{\mathrm{eff}} = \lambda \frac{2}{\sqrt{Pe^2 + 4} - Pe}, \tag{S15}$$

with $\lambda = \sqrt{\varepsilon_0 \varepsilon_l D/k}$ and $Pe = U\lambda/D$. Apparently, the structure of the diffuse layer is similar to the case without wall-normal flow, but the characteristic thickness is increased. There are two distinct regimes for the effective Debye length $\lambda_{\mathrm{eff}}$. For $Pe \ll 1$ it is unaffected and equal to $\lambda$, while for $Pe \gg 1$ it scales as $Pe\lambda$.

*Full analytical model* – To obtain the full analytical model, we solve the linearized equation S9 with the effective Debye length from equation S15 for the potential drop across the diffuse layer at the contact line:

$$\phi_{\mathrm{CL}}(V_l, Pe) = \frac{1}{2}(K - V + 1) - \sqrt{\frac{1}{4}\left(K - V\frac{\lambda_{\mathrm{eff}}}{\lambda} + 1\right)^2 + V\frac{\lambda_{\mathrm{eff}}}{\lambda}(K + 1) + KC\frac{\lambda_{\mathrm{eff}}}{\lambda}}. \tag{S16}$$

Finally, we use equation S11 along with equation S15 to find the surface charge density at the contact line:

$$\sigma_{\mathrm{CL}}(V_l, Pe, \theta) = \frac{\varepsilon_0 \varepsilon_l \phi_{\mathrm{CL}}}{\lambda_{\mathrm{eff}} g(\theta)} + \frac{\varepsilon_0 \varepsilon_s V_l}{d}. \tag{S17}$$

Equations S16 and S17 are the generalized forms of equations 6 in the main text for arbitrary liquid potentials $V_l$.

## §2.2 Underlying assumptions

The analytical model is built on a number of simplifying assumptions. We assume chemical equilibrium and an ideal solution, where the surface chemistry is only affected by deprotonation (equation S2), which is valid for pH $\gg$ pzc. We further apply the Debye-Hückel approximation $\phi < \phi_T$ and neglect the Stern layer capacitance. We assume no mutual influence between contact angle and flow effects and express contact angle effects with a term that is linearized around $\theta = \pi/2$. We assume that the potential drop across the diffuse layer is unaffected by the contact angle. For the flow effects, we assume a strictly wall-normal flow with velocity $U$ near the contact line and a quasi-one-dimensional structure of the diffuse layer.

## §2.3 Justification of contact angle coefficient

Our analytical model incorporates equation S10, as reported by Dörr and Hardt [S13]. It contains potentials and surface charge densities far from and at the contact line. For the model we assume that $\phi_{CL} = \phi_0 = \zeta$ in the absence of flow and use $g(\theta)$ to express deviations in the surface charge density close to the contact line. To test this assumption, we use our numerical simulations and plot the potential along the solid-liquid interface for different contact angles and negligible Péclet numbers. While there are some changes in the potential close to the contact line, they are smaller than $g(\theta)$. We thus deem our initial assumption justified.

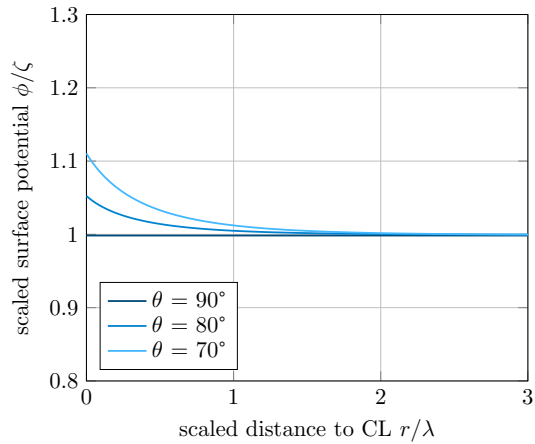

FIG. S4. Surface potential along the solid-liquid interface for different contact angles calculated by numerical simulations at zeta potential $\zeta = -50$ mV, Debye length $\lambda = 100$ nm and without contact line motion. The surface potential is scaled by the zeta potential and the distance to the contact line by the Debye length.

## §2.4 Justification of small Debye length gradient near the contact line

To analytically quantify the effects of the wall normal flow close to the contact line, we solve the one-dimensional version of equation S13. This approximation is only valid if the Debye length changes slowly enough as a function of distance from the contact line. Right at the gas-liquid interface, the boundary condition of zero normal electric field forces the isopotential contours to be normal to the interface, and thus for $\theta = \pi/2$, the local gradient in effective Debye length vanishes even at high Péclet numbers (Fig. S5). This result corroborates the validity of the 1D model. At higher $Pe$ the analytical one-dimensional model overestimates the expansion of the diffuse layer while predicted trends remain correct.

$^*$ hardt@nmf.tu-darmstadt.de

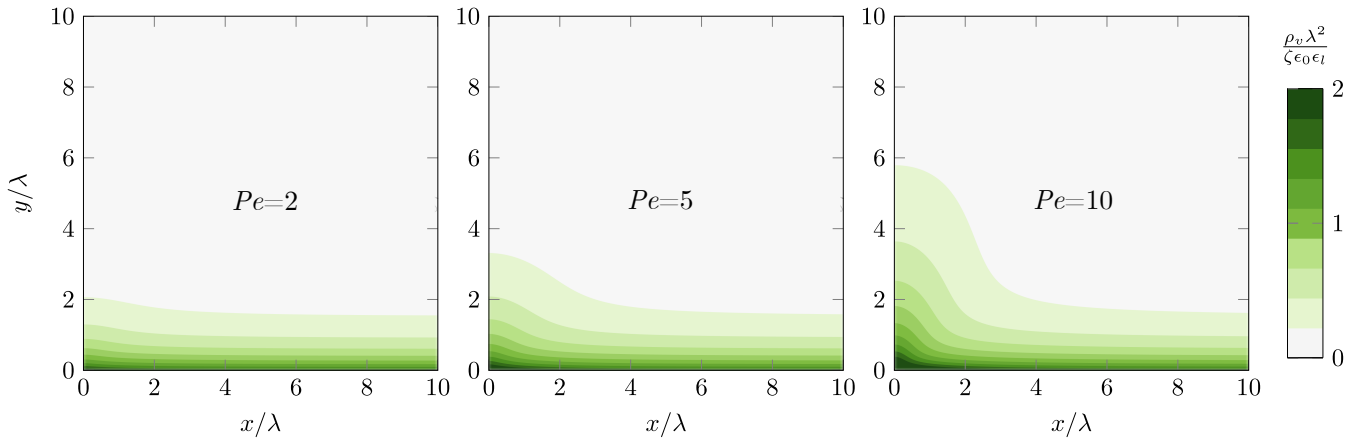

FIG. S5. Simulation results of the diffuse layer structure close to the contact line, located at $(0,0)$, for $\theta = \pi/2$ and different Péclet numbers. The coordinate axis are scaled with the Debye length $\lambda$ and the colors represent the space charge $\rho_v$ scaled with $\zeta \varepsilon_0 \varepsilon_l / \lambda^2$.



\end{document}